# Isotopes Tell Sun's Origin and Operation

O. Manuel[1], Sumeet A. Kamat[2], and Michael Mozina[3]

[1]*Nuclear Chemistry, University of Missouri, Rolla, MO 65401, USA*
[2]*Computer Science, University of Missouri, Rolla, MO 65401, USA*
[3]*Emerging Technologies, P. O. Box 1539, Mt. Shasta, CA 96067, USA*

**Abstract.** Modern versions of Aston's mass spectrometer enable measurements of two quantities – isotope abundances and masses – that tell the Sun's origin and operation. Isotope analyses of meteorites, the Earth, Moon, Mars, Jupiter, the solar wind, and solar flares over the past 45 years indicate that fresh, poorly-mixed, supernova debris formed the solar system. The iron-rich Sun formed on the collapsed supernova core and now itself acts as a magnetic plasma diffuser, as did the precursor star, separating ions by mass. This process covers the solar surface with lightweight elements and with the lighter isotopes of each element. Running difference imaging provides supporting evidence of a rigid, iron-rich structure below the Sun's fluid outer layer of lightweight elements. Mass measurements of all 2,850 known nuclides expose repulsive interactions between neutrons that trigger neutron-emission at the solar core, followed by neutron-decay and a series of reactions that collectively generate solar luminosity, solar neutrinos, the carrier gas for solar mass separation, and an outpouring of solar-wind hydrogen from the solar surface. Neutron-emission and neutron-decay generate ≈ 65% of solar luminosity; H-fusion ≈ 35%, and ≈ 1% of the neutron-decay product survives to depart as solar-wind hydrogen. The energy source for the Sun and other ordinary stars seems to be neutron-emission and neutron-decay, with partial fusion of the decay product, rather than simple fusion of hydrogen into helium or heavier elements.

## INTRODUCTION

The Apollo mission returned from the Moon in 1969 with soil samples whose surfaces were loaded with elements implanted by the solar wind. Analyses revealed neon of light atomic weight in these samples [1]. Fifty-six years earlier Aston [2] obtained lightweight neon by using diffusion to *mass fractionate* (*sort by mass*) neon atoms, in experiments that led to the discovery of isotopes and the development of the mass spectrometer [3]. The significance of lightweight neon in the solar wind could not be deciphered in 1969, when isotopic anomalies from stellar nuclear reactions and mass fractionation were not resolved, decay products of only two extinct nuclides had been found in meteorites [4, 5], and it was still widely believed that the solar system formed out of a well-mixed interstellar cloud with the composition of the Sun's surface.

A few observations that forced us to unpopular conclusions about the Sun's origin and operation will be shown. These data will be helpful to those seeking other explanations for the findings we encountered on this adventure. <u>Readers are encouraged to look at Figure 9 (p. 9) or Table 1 (p. 17) if they feel confused by the story connecting four decades of complicated experimental data to the few simple conclusions reached here</u>.

## ISOTOPE ABUNDANCE MEASUREMENTS

### Decay Products From Extinct, Short-Lived Isotopes

Mass spectrometric analyses of xenon isotopes first revealed the decay products of extinct $^{129}$I ($t_{1/2}$ = 16 Myr) [4] and $^{244}$Pu ($t_{1/2}$ = 80 Myr) [5] in meteorites. Decay products from these two extinct nuclides were soon found inside the Earth [6]. Fowler *et al.* [7] pointed out that the observed levels of short-lived radioactivity in the early solar system left little time for galactic mixing after the end of nucleosynthesis, i.e., the levels of short-lived radioactivity were higher than expected if the solar system formed from a typical interstellar cloud, representing an average sample of the entire galaxy. The discrepancy between isotope measurements and the nebular model for formation of the solar system grew larger as the decay products of even shorter-lived nuclides were discovered in meteorites.

Well-established chronometers of the early solar system presently include, in order of decreasing half-lives, $^{244}$Pu ($t_{1/2}$ = 80 Myr) [5], $^{129}$I ($t_{1/2}$ = 16 Myr) [4], $^{182}$Hf ($t_{1/2}$ = 9 Myr) [8], $^{107}$Pd ($t_{1/2}$ = 6.5 Myr) [9], $^{53}$Mn ($t_{1/2}$ = 3.7 Myr) [10], $^{60}$Fe ($t_{1/2}$ = 1.5 Myr) [11], $^{26}$Al ($t_{1/2}$ = 0.7 Myr) [12], and $^{41}$Ca ($t_{1/2}$ = 0.1 Myr) [13]. Most of these might have been produced in a supernova, and two of them, $^{244}$Pu and $^{60}$Fe, could only be made in a supernova explosion.

Rapid neutron-capture in a supernova explosion, the r-process, is believed to be the source of all nuclides heavier than $^{209}$Bi [14]. This includes extinct $^{244}$Pu and other actinide nuclides like $^{253}$U and $^{238}$U that are still alive and decaying to separate Pb isotopes. These form the basis for U/Pb age dating [15]. Spontaneous fission of $^{244}$Pu generates a distinctive pattern of Xe isotopes [16]. By combining the U/Pb and Pu/Xe age dating methods, Kuroda and Myers [17] were able to show that the most primitive meteorites started to retain Xe from the spontaneous fission of $^{244}$Pu soon after a supernova explosion occurred about 5 Gyr ago. Their results are shown in Figure 1.

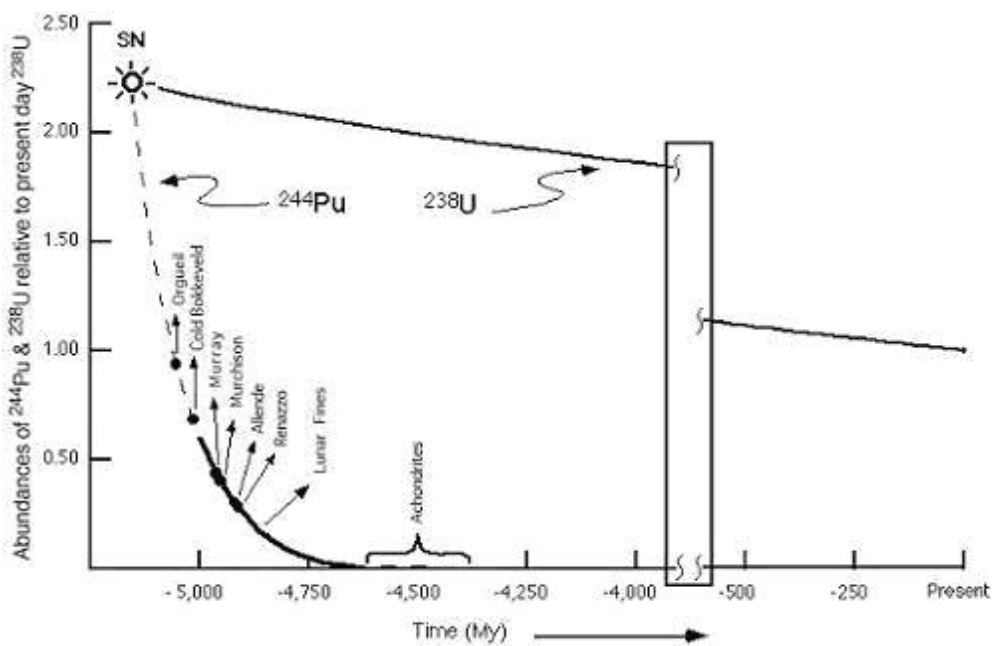

**FIGURE 1.** Combined U/Pb and Pu/Xe age dating shows a supernova exploded ≈ 5 Gy ago at the birth of the solar system [17].

Gaseous Xe isotopes from the fission of $^{244}$Pu were not retained quickly in the hot aftermath of the supernova. However Al/Mg age dating of refractory grains of silicon carbide and graphite demonstrates that some of these grains started to accumulate radiogenic $^{26}$Mg from the decay of $^{26}$Al within the first 1-10 My of the supernova explosion. Kuroda and Myers [18] note that the same supernova event produced short-lived $^{26}$Al ($t_{½}$ = 0.7 My), as well as longer-lived $^{244}$Pu ($t_{½}$ = 80 My). At the time of the supernova explosion, the *r*-process generated $^{244}$Pu and other trans-bismuth nuclides by rapid neutron capture, and the *x*-process made $^{26}$Al and $^{27}$Al at the surface of the supernova by spallation reactions [14, 18].

Radiogenic $^{26}$Mg, the decay product of extinct $^{26}$Al, is used to monitor the amount of $^{26}$Al initially trapped in meteorite minerals [19]. Those measurements show that values of the $^{26}$Al/$^{27}$Al ratio trapped in the earliest grains of silicon carbide correlate with the size of the particles [18]. In this sense the properties of large silicon carbide grains from the Murchison carbonaceous meteorite [19] mimic the properties of "fall-out" grains produced after the explosion of a nuclear weapon in air [18]. The first SiC grains that formed after the explosion trapped higher levels of extinct $^{26}$Al and grew larger; the smaller SiC grains started to form later and trapped aluminum with lower $^{26}$Al/$^{27}$Al ratios. The Al/Mg ages of these grains and their sizes are shown in Figure 2.

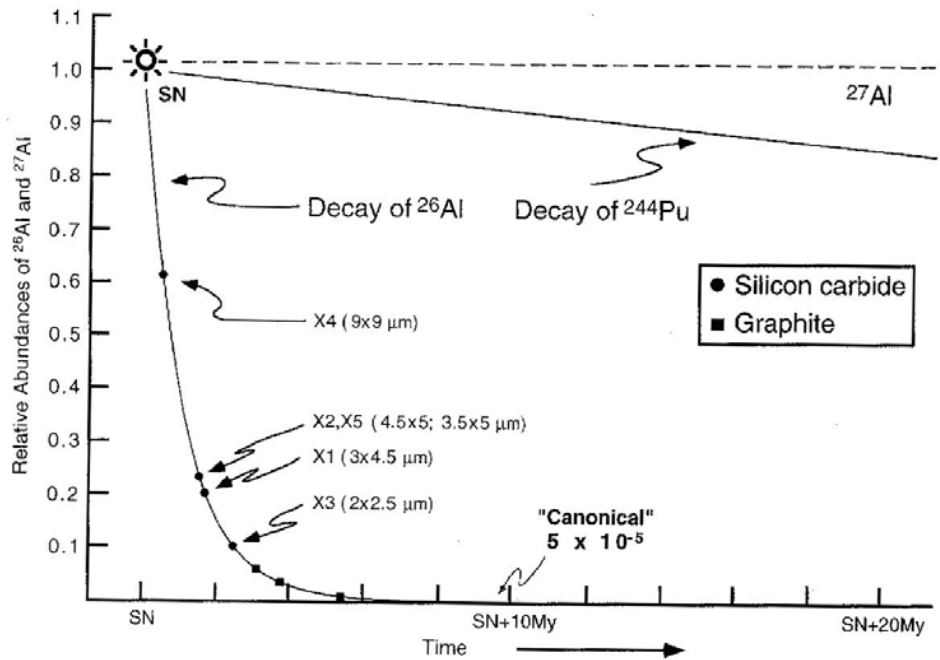

**FIGURE 2.** Kuroda and Myers [18] used $^{26}$Al/$^{26}$Mg age dating to show that refractory grains of silicon carbide from the Murchison meteorite [19] started to form within 1-10 My of the supernova explosion. The physical size of the grains and the relative amounts of radioactivity trapped in them mimic the "fall-out" grains produced after a nuclear weapons explosion in the atmosphere [18]. The steep slope of the solid line on the left represents the decay of $^{26}$Al. The other solid line shows the decay of $^{244}$Pu, about 100 times slower. The dashed horizontal line across the top of the figure represents stable $^{27}$Al.

## Strange Isotope Abundances From Stellar Nuclear Reactions Plus Mass Fractionation

Radiogenic $^{129}$Xe from the decay of extinct $^{129}$I ($t_{1/2}$ = 16 Myr) [4] and fissiogenic $^{131-136}$Xe from the spontaneous fission of $^{244}$Pu ($t_{1/2}$ = 80 Myr) [5] provided the first conclusive evidence that short-lived radioactive nuclides had been trapped in meteorites. Mass spectrometric analyses of xenon also revealed the first hint of a "strange" abundance pattern across all nine, stable isotopes of xenon [20]. Subsequent analyses confirmed this non-terrestrial abundance pattern for the xenon isotopes in other carbonaceous chondrites, and the term **AVCC** Xe was widely adopted to represent the xenon in **AV**erage **C**arbonaceous **C**hondrites [21]. The origin of the unusual isotope abundances in xenon and other elements proved to be extremely difficult to decipher because:

a.) Isotopes made by various stellar nuclear reactions were assumed to have been thoroughly mixed before meteorites formed and should therefore not be seen in meteorites as excesses or deficits of specific isotopes;
b.) Isotopic anomalies in meteorites from stellar nucleosynthesis reactions were found mysteriously embedded in elements that had been severely mass fractionated by some unknown process.

The possibility of isotope abundance changes from nuclear reactions and/or mass-dependent fractionation was acknowledged in Reynolds' first report of differences between the isotopic compositions of primordial xenon in meteorites and that in air [20]. Thus Reynolds suggested that *"The xenon in meteorites may have been augmented by nuclear processes between the time it was separated from the xenon now on earth and the time the meteorites were formed"*, and later he noted that *"On the other hand a strong mass-dependent fractionation may be responsible for most of the anomalies"* [ref. 20, p. 354].

Subsequent analysis of xenon isotopes released by stepwise heating of the carbonaceous meteorite, Renazzo, revealed a large enrichment of heavy xenon isotopes [22]. It was suggested that these heavy xenon isotopes might be from fission of extinct $^{244}$Pu [22], from a "carrier" of heavy xenon adsorbed on carbonaceous material [23], from mass-dependent fractionation [24], or from fission of super-heavy elements [25-27].

Then in 1972 it was reported [28] that the xenon released at about 600-1000°C from various carbonaceous meteorites is strongly enriched in both the lightest isotope, $^{124}$Xe, and in the heaviest isotope, $^{136}$Xe. This anomaly pattern could not be explained by any of the mechanisms proposed earlier [22-27]. The overall isotope anomaly pattern of this "strange" xenon is illustrated in Figure 3 for the xenon extracted from the 3CS4 mineral separate of the Allende meteorite [29].

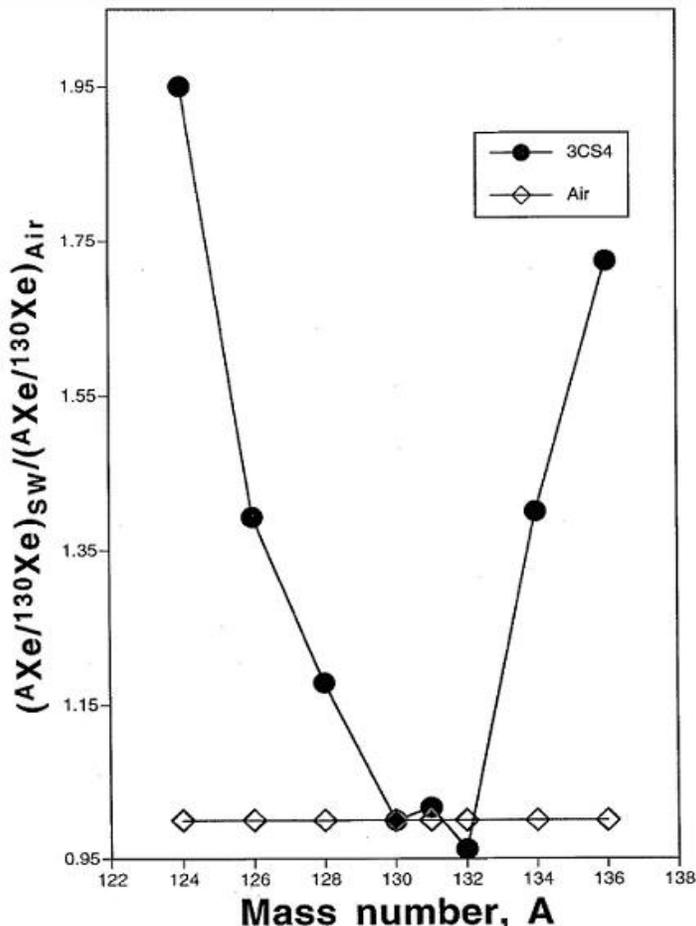

**FIGURE 3.** Relative to xenon in air, the xenon isotopes in mineral separate 3CS4 from the Allende meteorite [29] show large excesses of the neutron-poor and neutron-rich isotopes made by the p- and r-processes in a supernova explosion [14].

In Figure 3 xenon isotope abundances are shown relative to those in air, normalized to the middle isotope, $^{130}$Xe. Large isotopic anomalies are at the most neutron-poor and the most neutron-rich stable isotopes of xenon, $^{124}$Xe and $^{136}$Xe. They were made by violent nuclear reactions in a supernova explosion [14]. The p-process made $^{124}$Xe and the r-process made $^{136}$Xe. Neighboring stable xenon isotopes, like $^{126}$Xe and $^{134}$Xe, were also made at that time [14]. The middle isotopes, $^{128-132}$Xe, display smaller anomalies. The s-process [14] made these isotopes before the star reached the terminal supernova stage. The W-shaped isotopic anomaly pattern [29] in Figure 3 for xenon would be found later in other heavy elements with an extra component of r- and p-products (See p. 7, Figure 7).

In 1972 it was noted that differences in xenon isotope abundances *"cannot be explained by the occurrence of nuclear or fractionation processes that occurred within these meteorites"* [ref. 28, p. 99]. However excess heavy isotopes of argon and krypton were later found [29-31] in the meteorite sites that trapped "strange" xenon. A mass-dependent process outside of meteorites had altered the isotopes of argon, krypton and xenon in the Earth, the Sun, and meteorites. This mysterious mix of isotopic anomalies from nuclear-plus-fractionation would confuse and hamper efforts to decipher the stable isotope record of the early solar system over the next few decades.

Figure 4 shows the link [28] of excess $^{124}$Xe with excess $^{136}$Xe in the gas released from carbonaceous meteorites at ≈ 600-1000°C. **Xe-1** and **Xe-2** later proved to be characteristic of two distinct types of planetary noble gases in the **inner** and **outer** parts of the early solar system [32]: **P-1** ("normal" Xe-1, Kr-1, and Ar-1 only) and **P-2** ("strange" Xe-2, Kr-2, Ar-2, "normal" He and Ne). "Strange" Xe (**Xe-2**) in the upper right side of Figure 4 is enriched in the isotopes that were made in a supernova explosion by the r- and p-processes [14]. Mass-fractionation of "normal" Xe (**Xe-1**) along the dashed line in lower left side of Figure 4 could explain differences between xenon isotope abundances in the Earth, the Sun, and bulk meteorites, but *this would require about 9-stages of mass-dependent fractionation each operating at 100% efficiency*!

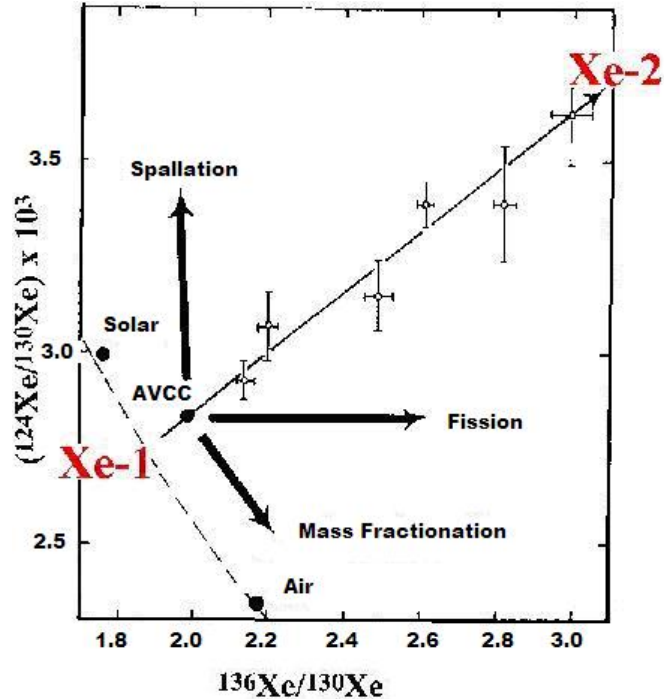

**FIGURE 4.** Meteorites trapped "normal" xenon (**Xe-1**) from the inner and "strange" xenon (**Xe-2**) from the outer parts of the solar system [28]. **Xe-2 is** enriched in $^{136}$Xe from the r-process and in $^{124}$Xe from the p-process [14]. Mass fractionation (dashed line) of "normal" xenon (**Xe-1**) relates **SOLAR** xenon with that in Earth's **AIR**. Bulk xenon in **A**verage **C**arbonaceous **C**hondrites (**AVCC**) is shifted away from this dashed fractionation line by the presence of "strange" xenon (**Xe-2**).

Similar mass fractionation effects had been seen in neon isotopes of the Fayetteville meteorite in 1967 [33]. Two years later, Marti [34] discovered solar-type xenon in the Pesyanoe meteorite and noted that isotope abundances in solar-type xenon and those in the terrestrial atmosphere might *". . . be related to each other by a strong mass-fractionating process"* [ref. 34, p. 1265]. The following year (1970) a common mass fractionation was reported across the isotopes of neon and xenon in air, in meteorites, and in lunar samples [24].

Despite a 1971 report of a mass-dependent covariance in the helium, neon, and argon isotopes in meteorites [35], *the site for multi-stage mass fractionation was unknown.* In the exciting early days of the space age, the possibility of isotope anomalies from mass fractionation seemed mundane compared to the exciting discovery that xenon isotopes in meteorites retained a record of the stellar nuclear reactions that produced them. An important correlation was thus overlooked: **SOLAR** and **AVCC** Xe are rich in lightweight isotopes (along the dashed fractionation line in Figure 4); the solar wind and carbonaceous chondrites are also rich in lightweight elements.

*The role of mass fractionation received little attention for the next three decades* as mass spectrometric analyses revealed large variations in the isotope abundances of several elements in meteorites. Variations in the atomic weights of neon in meteorites were, for example, identified as distinct isotopic components and labeled alphabetically: Ne-A, Ne-B, Ne-C, Ne-D, Ne-E, etc. [36-41].

Then in 1980 it was noted [42] that distinct neon components in meteorites, and differences between the isotopic compositions of bulk neon in air, in the solar wind and in meteorites, could be explained by the same mechanism that explained variations in neon isotope abundances in the Fayetteville meteorite [33]. The dashed line in Figure 5 shows the mass-dependent fractionation reported across neon isotopes in the Fayetteville meteorite. The 1980 paper showed that *a mix of mass-fractionated neon with cosmogenic neon could also explain <u>all</u> members of the neon alphabet that had been identified in meteorites* [36-41].

Figure 6 illustrates this mass-dependent relationship as the area bounded by curved lines when planetary Ne-A is fractionated. Within this area are (in order of decreasing $^{20}$Ne abundance) Ne-D, Ne-S (solar), Ne-B, Air, Ne-A1, M-800, M-1200 and M-1400 (neon released from the Murchison meteorite at 800° C, 1200° C and 1400° C), and the original upper limit on Ne-E. Cosmogenic Ne lies to the right of Figure 6 at $^{20}$Ne/$^{22}$Ne ≈ $^{21}$Ne/$^{22}$Ne ≈ 1.0. The addition of cosmogenic neon to fractionated planetary neon, as shown by the bold line connecting Ne-A1 with Ne-A2, explained all the neon isotope data available in 1980 [36-41], including Ne-C, M-1000, Ne-A2, M-1100 [42].

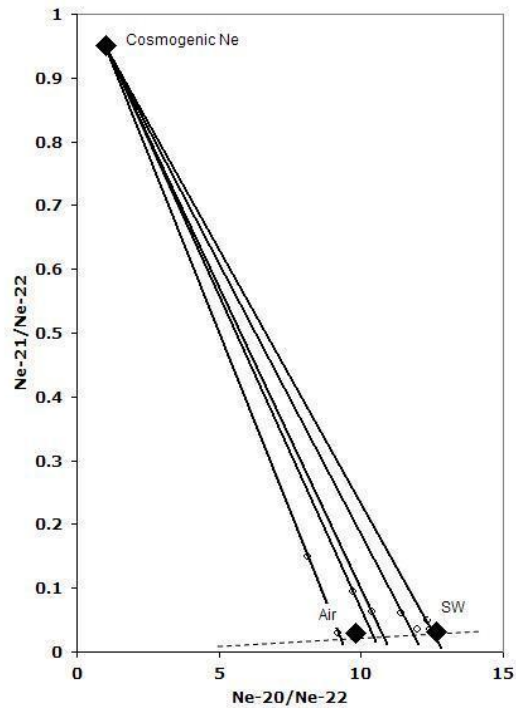

**FIGURE 5.** In 1967 neon isotopes in air, in the solar wind (SW), and those released by stepwise heating of dark portions of the Fayetteville meteorite were reported to be a mix of cosmogenic neon made by spallation reactions (top, left) with mass-fractionated primordial neon lying along the dashed line [33]. The large filled diamonds identify **Cosmogenic**, **Air**, and **SW** (solar-wind) neon. The small open circles show neon released by stepwise heating of the Fayetteville meteorite [33].

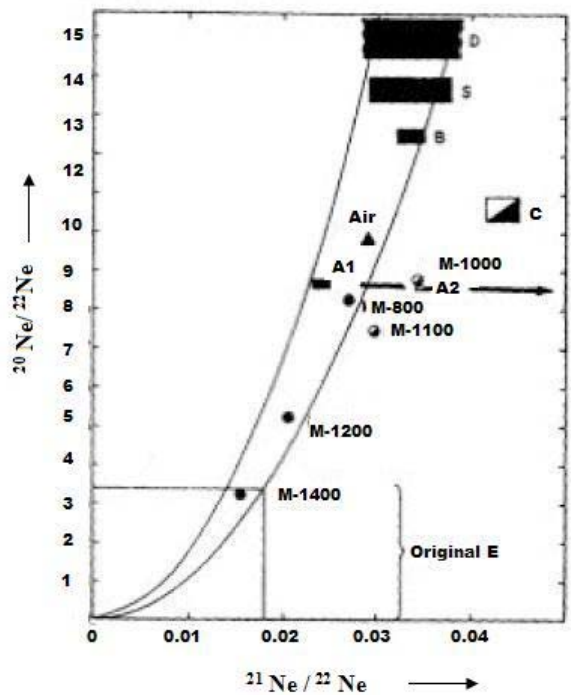

**FIGURE 6.** In 1980 it was shown that the neon isotopes in **Air**, in the solar wind (**S**), and in various parts of meteorites [36-41] could be explained as simple mixes of mass-fractionated planetary neon with the cosmogenic neon that is produced when high-energy cosmic rays induce spallation reactions on heavier target nuclei in meteorites [42].

All neon components in meteorites [36-41] and the bulk neon in air, in the solar wind, and in meteorites lie along the dashed fractionation line in Figure 5, after correcting for neon made by cosmic-ray induced spallation reactions [42]. *The site of such severe fractionation was not recognized in 1980, although s-products from the interior of a star had been found with the most severely mass-fractionated form of neon, Ne-E* [43].

However a complex mix of nuclear-plus-fractionation effects had also been recognized in the isotopes of other elements in meteorites by 1980, including the refractory element, magnesium. Clayton and Mayeda at the University of Chicago [44] and Wasserburg and coworkers at Cal Tech [45] reported that a mass-dependent fractionation process had enriched the heavy isotopes of oxygen and magnesium in refractory inclusions of the Allende meteorite. The name, **FUN** anomalies, was assigned to the strange isotope abundances in meteorites that arise from a combination of **F**ractionation plus **U**nknown **N**uclear effects [45]. Figure 4 shows that evidence of **FUN**-like anomalies had been noted earlier in the isotopes of xenon in meteorites.

In heavier, trans-iron elements, excesses of the intermediate mass isotopes of Xe [43] and Te [46] were reported in some meteorite grains from the s-process of nucleosynthesis [14]. These isotope anomalies are complementary to the excesses of lightweight and heavy isotopes of Xe and Te [28-31, 46] reported in other meteorite grains from the p- and r-processes of nucleosynthesis [14]. In a 1980 review on isotopic anomalies in meteorites, Begemann [47] noted that these findings have overthrown the classical picture of the pre-solar nebula as a *". . . hot, well-mixed cloud of chemically and isotopically uniform composition"* [ref. 47, abstract, p. 1309].

The validity of Begemann's conclusion became more evident in the next decade, as complementary excesses (+) and deficits (-) of the same isotope were identified in several other trans-iron elements in meteorites. He and his associates at Mainz reported barium [48] and krypton [49] enriched in isotopes made by the s-process in the Murchison meteorite, and in the Allende meteorite tellurium [50] and xenon [51] isotopes made by the r-process and separated from their radioactive precursor isotopes within the first few hours after the supernova explosion.

The schematic drawing in Figure 7 illustrates that stellar debris formed "normal", as well as the "strange" mirror-image (+ and -) isotopic anomaly patterns that Begemann [52] identified before mixing of the isotopes of Ba, Nd and Sm in inclusions from the Allende (top) and Murchison (bottom) carbonaceous meteorites.

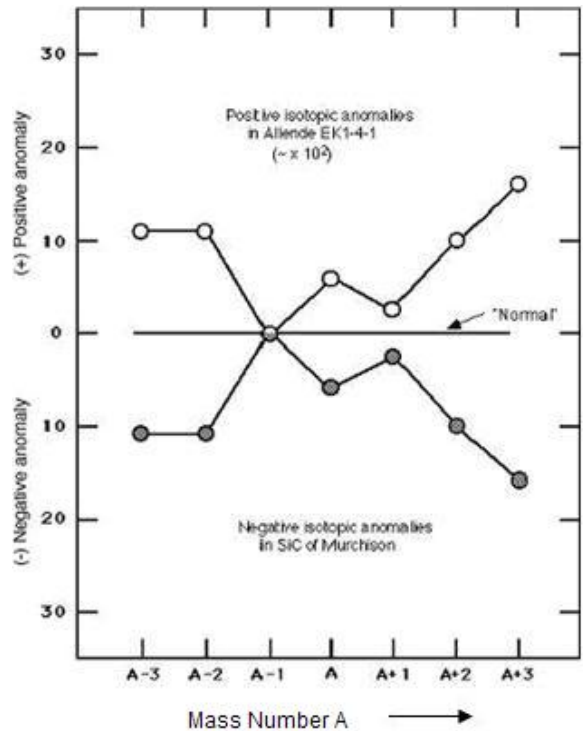

**FIGURE 7.** Barium, neodymium, and samarium in silicon carbide inclusions of the Murchison meteorite (bottom) are enriched in isotopes made by the s-process [14]. Isotopes of the same elements made by rapid nuclear reactions in the supernova, the r- and p-processes [14] are enriched in inclusion EK-1-4-1 of the Allende meteorite (top). The p-process makes the light isotopes, the s-process makes the intermediate ones, and the r-process makes the heavy isotopes of trans-iron elements [14]. Anomalies in the bottom half of this figure are about two orders-of-magnitude larger than the positive anomalies in the upper half of the figure.

This paper is concerned primarily with the origin and operation of the Sun, so we will not attempt to give a complete review of all reports of nucleogenetic isotopic anomalies in meteorites from leading research institutions around the world. Suffice it to say that these findings confirmed Begemann's 1980 conclusion that the classical picture of a homogeneous pre-solar nebula has been overthrown [47]. However, the possibility of a mistake in the classical picture of a hydrogen-filled Sun [46, 53, 54] continued to receive little attention.

Before leaving the subject, some surprising results from the University of Tokyo, Harvard, and Cal Tech should be mentioned. Careful analysis of iron meteorites at the University of Tokyo revealed a startling discovery: Massive iron meteorites retained isotopic anomalies from the stellar nuclear reactions that made the stable isotopes of molybdenum [55, 56]. These isotopes include $^{92}$Mo from the p-process, $^{96}$Mo from the s-process, $^{100}$Mo from the r-process, and other isotopes from a mix of nucleosynthesis reactions [57]. High precision mass spectrometry showed that these Mo isotopes never completely mixed after stellar nucleosynthesis, even in the massive objects thought to be highly differentiated iron meteorites. Recent analyses at Harvard [58] and Cal Tech [59] have confirmed Mo isotope anomalies in ordinary iron meteorites.

These findings confirm the suggestion [46, 53, 54] that iron meteorites, and the cores of the terrestrial planets, likely formed directly from iron-rich supernova debris, rather than by geochemical differentiation and extraction of iron from an interstellar cloud. Linked chemical and isotopic heterogeneities in meteorites, to be discussed below, offered the first compelling evidence that heterogeneous supernova debris formed the entire solar system, without being first injecting into the interstellar medium.

## The Link Between Chemical And Isotopic Heterogeneities

Figure 8 shows an unexpected finding from isotope analyses on mineral separates of the Allende meteorite in 1975 [29]: Primordial helium accompanied only "strange" Xe, not "normal" Xe, in the highly radioactive supernova debris that condensed to form the earliest meteorite grains.

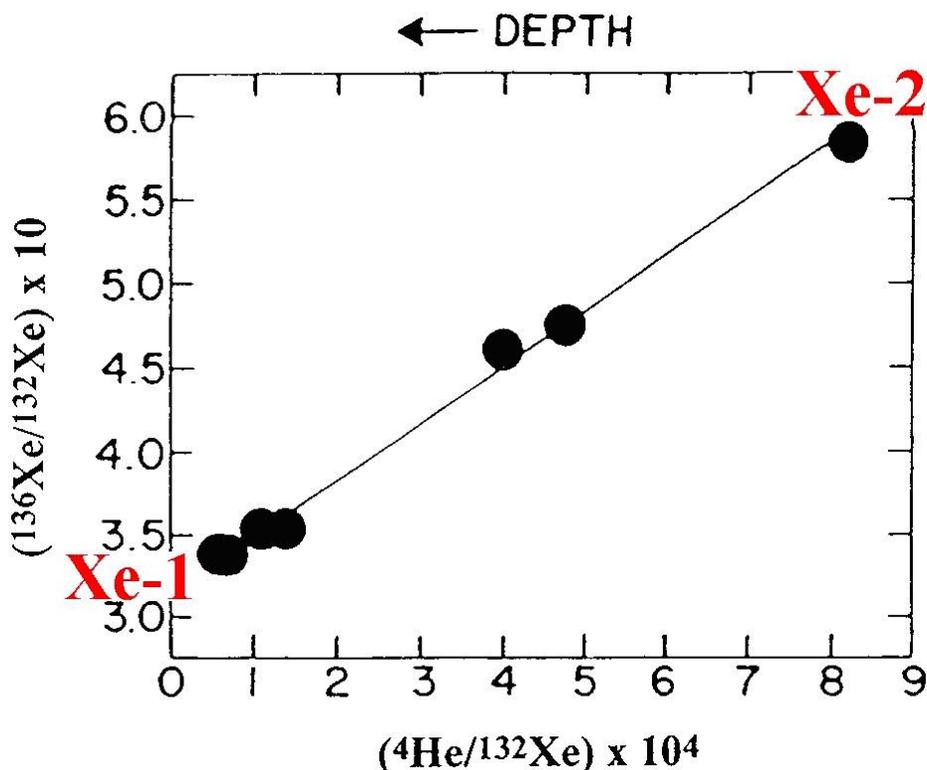

**FIGURE 8.** Primordial helium accompanied "strange" xenon (**Xe-2**), not "normal" xenon (**Xe-1**), when minerals of the Allende meteorite formed [53, 54]. This link, between the abundance of a light element, helium, with the abundance of an isotope in a heavy element, $^{136}$Xe in xenon, was one of the first clues that the r- and p-processes made **Xe-2** in the He-rich outer regions of the supernova that produced the solar system and **Xe-1** came from its iron-rich interior. In Figure 8 the coefficient of correlation between $^4$He and excess $^{136}$Xe in mineral separates of the primitive Allende carbonaceous meteorite is > 99%. Stellar depth and the disappearance of helium by fusion increase from right to left, as shown by the arrow at the top of this figure.

The element/isotope correlation shown in Figure 8 and the link of primordial helium and neon with certain other isotopes of xenon, krypton, and argon were the first compelling evidence that *the solar system emanated from highly radioactive material that was heterogeneous in the abundances of light elements, as well as in the isotopes of heavy elements* [46, 53, 54]. The surprising link between the elemental abundance of primordial helium and excess $^{136}$Xe from the r-process of nucleosynthesis was the primary reason for proposing that the solar system formed directly from the heterogeneous debris of a supernova [53, 54], as shown below:

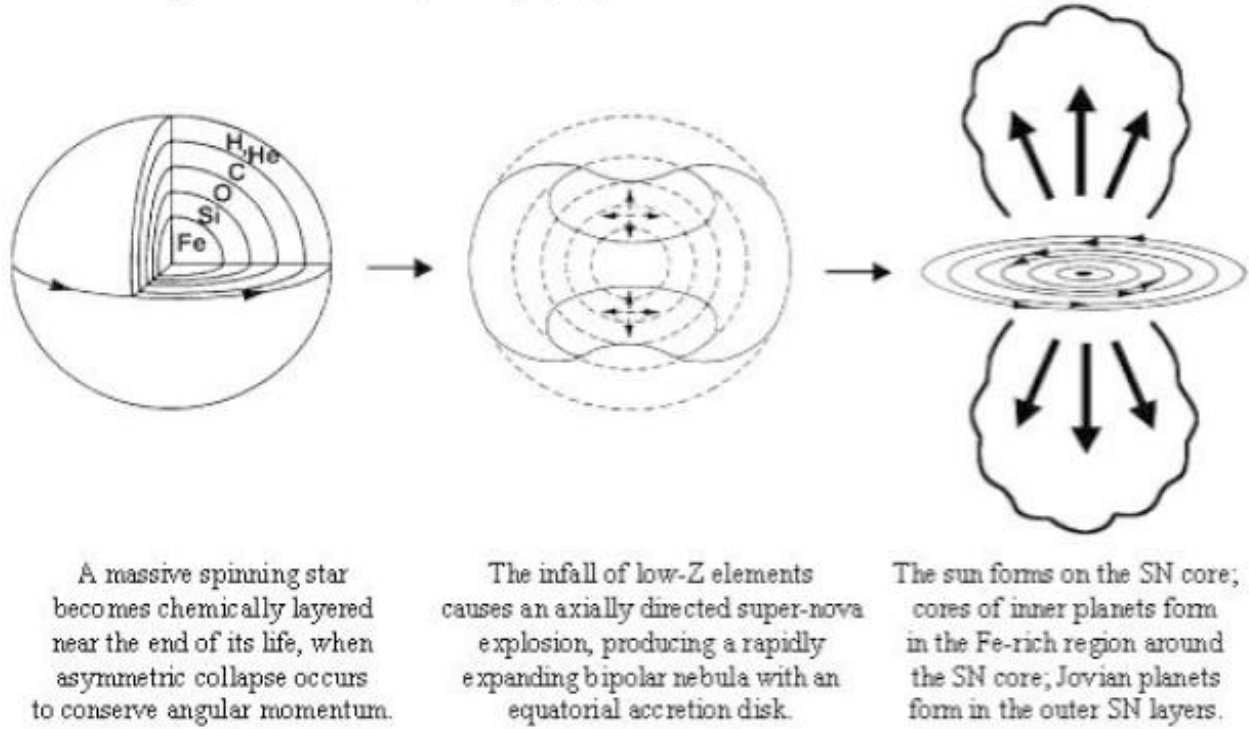

A massive spinning star becomes chemically layered near the end of its life, when asymmetric collapse occurs to conserve angular momentum.

The infall of low-Z elements causes an axially directed super-nova explosion, producing a rapidly expanding bipolar nebula with an equatorial accretion disk.

The sun forms on the SN core; cores of inner planets form in the Fe-rich region around the SN core; Jovian planets form in the outer SN layers.

**FIGURE 9.** The link of primordial helium with "strange" xenon, and its absence from the noble gas component with "normal" xenon, was interpreted as evidence that the solar system formed directly from heterogeneous supernova debris [53, 54]. "Strange" Xe (**Xe-2**) came from the He-rich outer part of the supernova. "Normal" xenon (**Xe-1**) came from the hot Fe-rich interior of the star, which was depleted of He and other lightweight elements [14]. Dr. Kiril Panov has correctly noted that a more radical evolutionary scheme may be required if our conclusions for the Sun are true for other stars in this galaxy and others.

According to Figure 9, the link of primordial helium with excess $^{136}$Xe was established in the He-rich outer part of the supernova when the r-process made $^{136}$Xe there [53, 54]. Figure 9 offers a viable explanation for the high levels of decay products from extinct $^{129}$I and $^{245}$Pu that had been discovered in meteorites and in the Earth [4-7]. The scenario in Figure 9 can also account for other nucleogenetic isotopic anomalies [43-52, 55-59] that would be found later in meteorites and the decay products of other extinct isotopes [8-13], including the presence of live $^{60}$Fe from the supernova core *"in the cradle of the solar system"* [ref. 60, p. 1116].

Subsequent measurements confirmed that "strange" xenon (**Xe-2**) was linked with the lightweight elements that formed gaseous planets [61, 62] and diamond and graphite inclusions of meteorites [30-32] in the outer region of the solar system [46, 53, 54]. "Normal" xenon (**Xe-1**) was linked with the iron, nickel and sulfur that formed troilite inclusions in meteorites [63], rocky planets [64] and the Sun [65] in the inner solar system.

Later meteorite analyses revealed nucleogenetic isotopic anomalies in other elements linked with the chemical composition of the carrier phase. Although the connection was not as clear-cut as the either/or association of primordial helium and neon with "strange" xenon, krypton and argon [32], these associations also extended across planetary distances to the sites where different classes of planets and meteorites formed.

In 1973 it was reported that carbonaceous meteorites trapped a primitive nuclear component of *"almost pure $^{16}O$"* [ref. 66, p. 485] when they formed. In 1976 it was found that the excess $^{16}$O is characteristic of six distinct classes of meteorites and planets, *"none of which can be derived from another by fractionation processes alone"* [ref. 67, p. 17]. Figure 10 shows the levels of excess $^{16}$O in different types of meteorites [67].

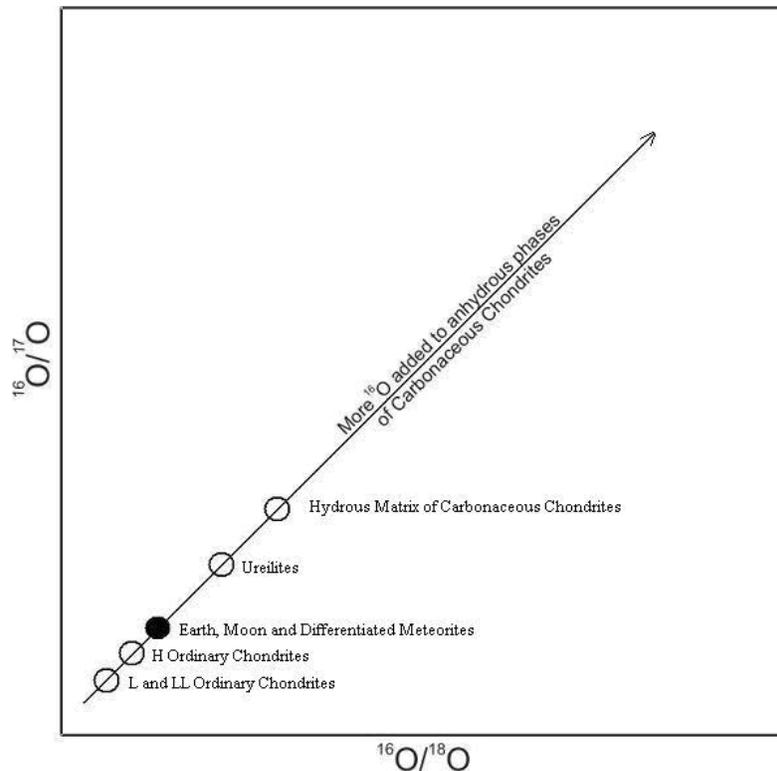

**FIGURE 10.** Different classes of meteorites and planets can be grouped into at least six different categories based on the levels of $^{16}O$ in their oxygen [67]. The sixth and most $^{16}O$-rich category is the anhydrous phase of carbonaceous chondrites. This association and the presence of excess $^{16}O$ in the Sun itself [68] are unexplained by tiny injections of "alien" material [69-71].

Levels of mono-isotopic $^{16}O$, like those of "strange" xenon, varied across planetary distances, closely linked with chemical heterogeneities in the material that formed the different types of meteorites and planets. Fowler [69], Cameron [70], and Wasserburg [71] correctly noted that many isotopic anomalies from the decay of extinct isotopes or nucleosynthesis involve only a tiny fraction ($\approx 10^{-5}$ to $10^{-4}$) of the material in the solar system. The data in Figures 8 and 10 show that some of these isotopic anomalies were closely associated with elements that may comprise a larger fraction of the total mass of the solar system.

It would be surprising if the Sun accreted none of the material with anomalous isotope abundances. Earlier this year scientists at Osaka University and the CNRS Research Facility reported that at least 2.0% ($\pm 0.4\%$) of the oxygen in the Sun is mono-isotopic $^{16}O$ [68]. Thus oxygen in the Sun would plot in the upper right side of Figure 10, with even more $^{16}O$ than the Earth, the Moon, ordinary meteorites, or the hydrous matrix of carbonaceous chondrites. The amount of excess $^{16}O$ reported in the Sun (>2%) is bracketed by the amounts of excess $^{16}O$ (<4%) observed in the anhydrous phase of carbonaceous chondrites [66, 67].

Xenon in the solar wind is mostly a mass-fractionated form of "normal" xenon (**Xe-1**), like that on Earth (See Figure 4). However it was noted in 1976 [65] that "strange" xenon (**Xe-2**) might account for about 7% of the $^{136}Xe$ in the Sun. Researchers at the University of Minnesota [72] recently estimated that "strange" xenon (**Xe-2**) accounts for about 8% of the $^{136}Xe$ in the Sun.

Lee *et al.* [73] noted other links between isotopic anomalies and the chemical composition of the carrier phase in meteorites. In addition to the "normal" isotope abundances seen in heavy elements here on Earth and in iron sulfide inclusions of meteorite from a mix of stellar nuclear reactions [14], specific nucleosynthesis reactions generated excess r- and p-isotopes in the heavy elements that became trapped in diamond inclusions of meteorites, and excess s-isotopes in the heavy elements that became trapped in the silicon carbide inclusions. Silicon carbide also trapped mass-fractionated neon isotopes [42] (See Ne-E in Figure 6), confirming the **FUN** link [44, 45].

Refractory diamond and silicon carbide grains maintained the record of linked chemical and isotopic variations in different classes of meteorites, e.g., *"in unmetamorphosed members of all seven chondrite classes"* [ref. 74, p. 115]. The linkage of r- and p-products with diamonds, and the linkage of s-products with silicon carbide, are so strong that Huss and Lewis [74] were able to estimate the amounts of each type of inclusion by measuring noble gas isotopes in acid residues of meteorites.

Thus the scenario shown in Figure 9 for the origin of the solar system explains the decay products of short-lived isotopes, nucleogenetic isotopic anomalies, and the links preserved between elemental and isotopic heterogeneities in meteorites and planets. *Four enigmas*, unexplained by Figure 9, will be addressed in the next sections:

- a.) Why is the chemical composition of the Sun's surface [75], as shown below in Figure 11, so unlike that expected in the central object left from the supernova explosion (Figure 9)?
- b.) Why are fractionation-plus-nuclear effects linked for the isotope anomalies in meteorites [28, 44, 45]?
- c.) Why is the composition of carbonaceous chondrites like that of the solar photosphere [75]?
- d.) What is the current source of luminosity in the Sun?

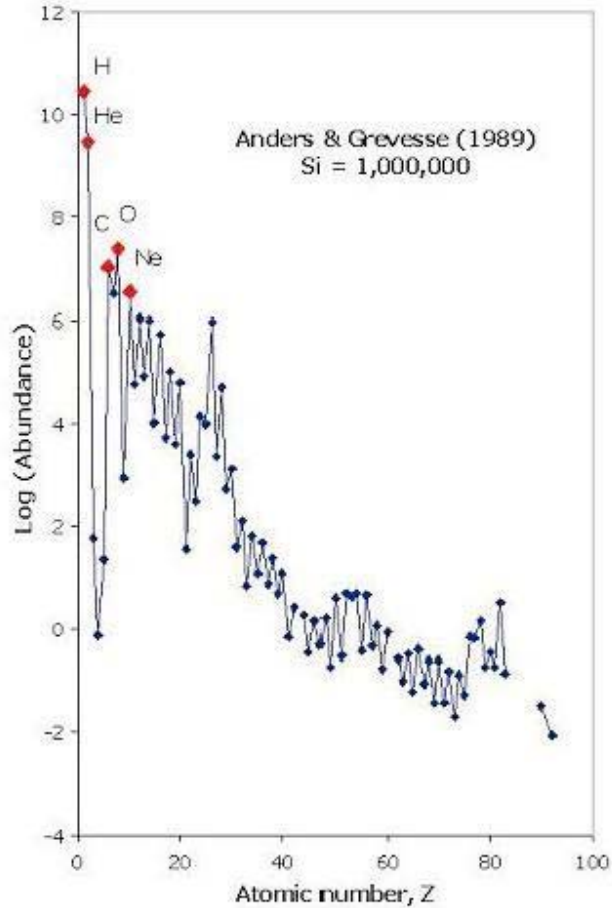

**FIGURE 11.** Lightweight elements are dominant in the Sun's photosphere [75]. Abundances of the heavier elements tend to decrease exponentially with increasing atomic number, Z.

## The Site Of Multi-Stage Mass-Dependent Fractionation

Other nucleogenetic isotope anomalies and the decay products of other extinct nuclides were found after the scenario in Figure 9 was proposed for the origin of the solar system [53, 54, 76]. It was suggested that these findings might be explained instead by the addition of a *small fraction* ($\approx 10^{-5}$ to $10^{-4}$) of "alien" nucleogenetic material [69-71] to the early solar system. Clayton [77] even suggested that the link of primordial helium with "strange" xenon (Figure 8) might indicate that both are alien nucleogenetic products, trapped in circumstellar carbon dust grains that migrated into the early solar system.

Manuel and Hwaung [78] took a different approach. *They assumed that the Sun is a mix of the components seen in meteorites* and used isotope abundances in the solar wind to estimate the fraction of each primitive component in the Sun. Since primordial helium and neon are only found with "strange" **Xe-2**, **Kr-2**, and **Ar2** in meteorites [32], they assumed that helium and neon in the Sun came from this source. This assumption is consistent with the mass-dependent relationship seen [35] across the isotopes of helium and neon (See also Figures 5 and 6).

*Manuel and Hwaung [78] found a ≈ 9-stage mass fractionation process*! About nine theoretical stages of mass-dependent fractionation have each enriched the abundance of the lighter mass (**L**) neon isotope relative to that of the heavier mass (**H**) one in the solar wind by the square root of (**H/L**). Although less well defined, the isotopes of solar-wind helium seem to have been sorted by the same mass-dependent process [78].

The 9-stage fractionation process extends across the isotopes of the heaviest noble gas, but solar-wind xenon is mostly a mass-fractionated form of "normal" xenon like that in air (**Xe-1** in Figures 4 and 8). Assuming that the same process sorts the intermediate noble gas isotopes, they showed [78] that solar krypton is a mix of the "normal" (**Kr-1**) and "strange" (**Kr-2**) seen in meteorites [32], while solar argon is the "strange" **Ar-2** [32] that accompanies primordial helium and neon in meteorites. Their results are shown in Figure 12.

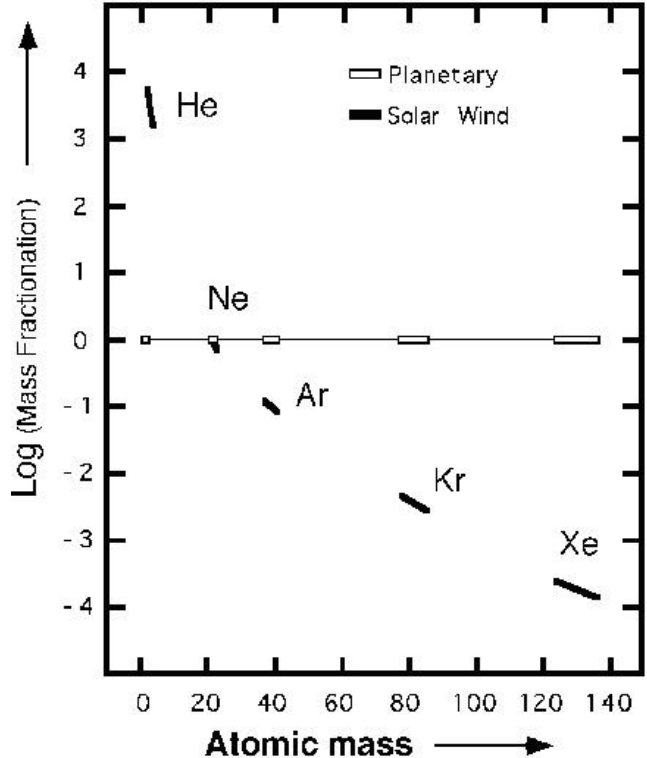

**FIGURE 12.** A common, 9-stage mass fractionation process selectively enriches light isotopes of He, Ne, Ar, Kr and Xe in the solar wind relative to those in planetary material [78]. Solar noble gases (in bold) are a mix of the two planetary components [32]: {He, Ne, Ar-2, Kr-2, and Xe-2}$_{P-2}$ + {Ar-1, Kr-1 and Xe-1}$_{P-1}$ = {**He, Ne, Ar-2, (Kr-2 + Kr-1) and Xe-1**}$_{SUN}$. [78].

If **H** and **L** are the masses of the heavier and lighter isotopes, respectively, then the empirical mass-dependent fractionation power law that Manuel and Hwaung [78] found to be common across the stable isotopes of the five noble gases in the solar wind, from A = 3 to 136 mass units, is

$$\log(MassFractionation) = 4.56\log(H/L) \tag{1}$$

When element abundances in the photosphere were corrected for this fractionation, the most abundant elements in the interior of the Sun were found to be Fe, Ni, O, Si, S, Mg and Ca [78]. These same elements comprise 99% of ordinary meteorites [79]. The probability, **P**, is essentially zero ($P < 2 \times 10^{-33}$ [80]) that this agreement between the composition of the Sun and that of ordinary meteorites is fortuitous.

Any remaining doubts about planetary systems forming directly from supernova debris were further reduced by Wolszczan's 1994 report [81] of rocky, Earth-like planets orbiting pulsar PSR 1257+12. Wolszczan's observations [81] and isotope measurements [78] on the solar wind thus explain *the first enigma*: Supernova debris formed the solar system but lightweight isotopes and elements [75] are abundant in the photosphere because *the Sun selectively moves lightweight elements and the lighter isotopes of each element to the solar surface*.

Isotope measurements also explain why heavy elements are more abundant in solar flares than in the solar wind: Flares are energetic events that by-pass about 3.4 of the 9-stages of mass fractionation [80]. The *Wind Spacecraft* also observed that heavy elements were systematically enriched, by several orders of magnitude, in material ejected from the interior of the Sun by an impulsive solar flare [82].

Abundances of s-process nuclides in the photosphere confirm mass fractionation in the Sun [83]. For nuclides made by slow-neutron capture [14], the steady-flow abundance, *N*, of nuclides of successive mass numbers, A-1, A, A+1, is expected to be inversely proportional to their neutron-capture cross sections, $\sigma$:

$$N(A-1)*\sigma(A-1) = N(A)*\sigma(A) = N(A+1)*\sigma(A+1) \tag{2}$$

This prediction [14] has been confirmed across the isotopes of samarium that were made by the s-process, $^{148}$Sm and $^{150}$Sm, and across those of tellurium, $^{122}$Te, $^{123}$Te and $^{124}$Te [83]. Over the total mass range of s-products in the photosphere, however, values of the $N*\sigma$ product decline by ≈ 5 orders of magnitude as the mass number, A, increases from 25 to 207 mass units. This is shown below in Figure 13.

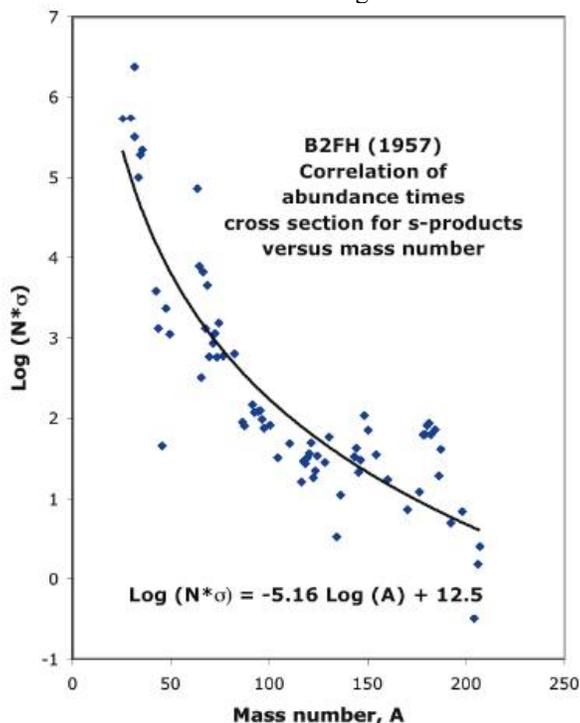

**FIGURE 13.** Values of log ($N*\sigma$) for the nuclides in the photosphere that B2FH [14] identified as s-products decline exponentially with increasing mass number, A. The 72 data points shown here [14] cover a mass range of 25-207 mass units.

The mass-fractionation relationship defined by the abundances of s-products in the photosphere (Figure 13) indicates that the lightweight ones have been enriched by ≈ 10-stages of mass-dependent fractionation. Despite the scatter of data points, correction of the solar abundance data of Anders and Grevesse [75] for this fractionation relationship also yields Fe, Ni, O, Si and S as the most abundant elements in the interior of the Sun [83].

This reinforces the validity of the answer given earlier for the *first enigma*, and it suggests a related answer for the *second enigma* and the *third enigma*:

a.) The chemical composition of the Sun's surface [75] (Figure 11) does ***not*** match that expected in the central object left from the supernova explosion (Figure 9) because *the Sun selectively moves lightweight elements and the lighter isotopes of each element to the solar surface*.

b.) Fractionation-plus-nuclear effects are linked in meteorites [28, 44, 45] because *mass-fractionation is a common occurrence in stars, including the one that gave birth to the solar system (Figure 9). Thus stellar nuclear reactions generated new isotopes in material that was already mass fractionated*.

c.) The composition of carbonaceous chondrites is like the photosphere [75] *because lightweight elements are now sorted into the photosphere, as they had earlier been sorted into that part of the parent star that would later form carbonaceous chondrites and gaseous planets*.

Thus solar mass fractionation explains: i) Excess lightweight isotopes in the solar wind (Figure 12), ii) Elevated abundances of heavy elements and heavy isotopes in solar flares [80, 82], iii) Mass fractionated s-products in the photosphere (Figure 13), and it also explains the second of the two serious difficulties that Nobel Laureate William A. Fowler [84] identified in the most basic concepts of nuclear physics:

a.) *"On square one the solar neutrino problem is still with us (...), indicating that we do not understand how our own star really works."*
b.) *"On square two we still cannot show in the laboratory and in theoretical calculations why the ratio of oxygen to carbon in the Sun and similar stars is close to two to one (...)"* [ref. 84, p. xi]

It was recently shown [83] that mass fractionation changes the oxygen to carbon ratio from O/C ≈ 10 in the Sun and similar stars to the surface value of O/C ≈ 2. The solar neutrino problem [84] will be addressed in the section on solar luminosity, after a brief look at visual images of the iron-rich structure below the Sun's fluid outer layers.

## Visual Images Of A Rigid Iron-Rich Solar Surface

Mozina [86] reported a surprising discovery in the spring of this year, *"While viewing images from SOHO's EIT program, I finally stumbled across the raw (unprocessed EIT images) marked "DIT" images that are stored in SOHO's daily archives. After downloading a number of these larger "DIT" (grey) files, including several "running difference" images, it became quite apparent that many of the finer details revealed in the raw EIT images are simply lost during the computer enhancement process that is used to create the more familiar EIT colorized images that are displayed on SOHO's website. That evening in April of 2005, all my beliefs about the sun changed."* Here are images, over a 60-hour (2.5-day) period on 27-29 May 2005, of the rigid structure below the fluid photosphere.

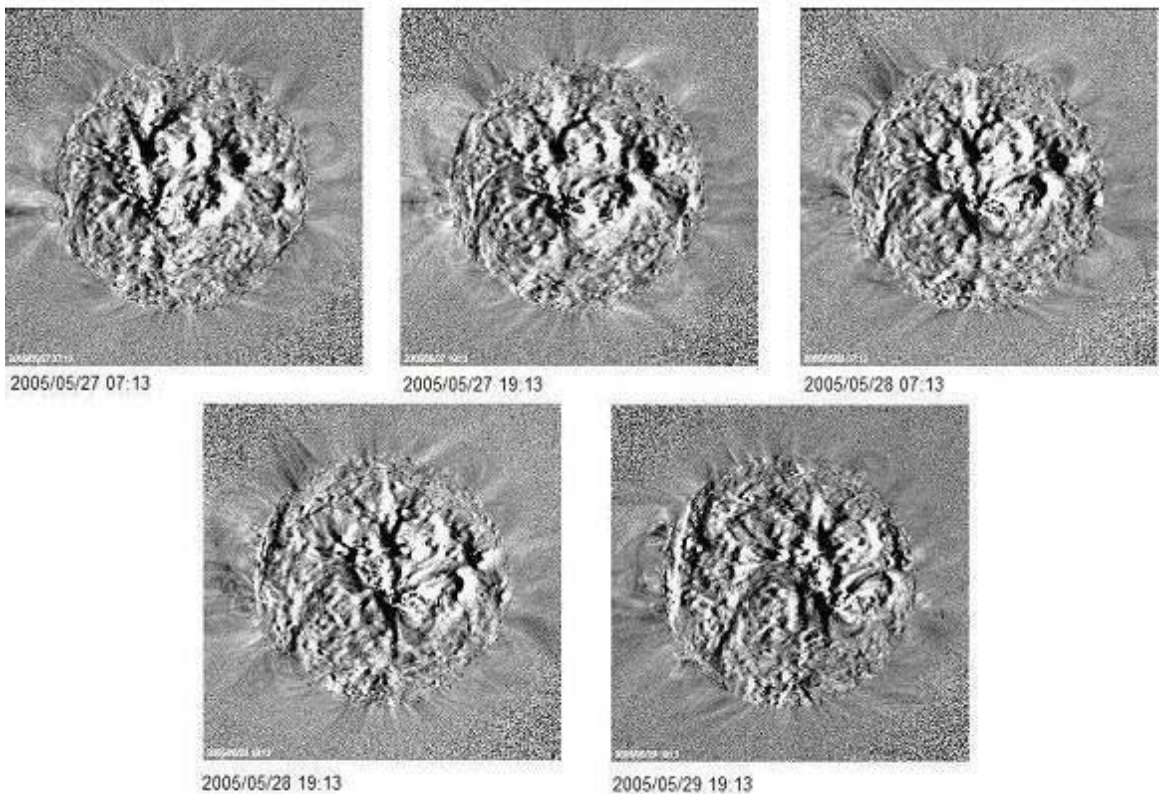

**FIGURE 14.** Running difference images from SOHO using a 195 Å filter to enhance light Fe (IX) and Fe (X) emissions. Surface features in the iron-rich surface below the Sun's fluid, hydrogen-rich photosphere are visible for days or weeks. From videos of these images, Mozina shows that this surface rotates uniformly, from pole to equator, every 27.3 days [86].

Mozina also noticed that images from the TRACE satellite revealed a solar flare and the release of volatiles from a small section of the sun's surface on 28 Aug. 2000. Figure 15 shows this region, using the 171Å filter that is specifically sensitive to the iron ion (Fe IX/X) emissions.

The first serious difficulty Fowler identified in nuclear astrophysics — *"the solar neutrino problem"* [Ref. 84, p. xi], and the <u>*fourth enigma*</u> — "*the current source of luminosity in the Sun*", will be addressed in the next section.

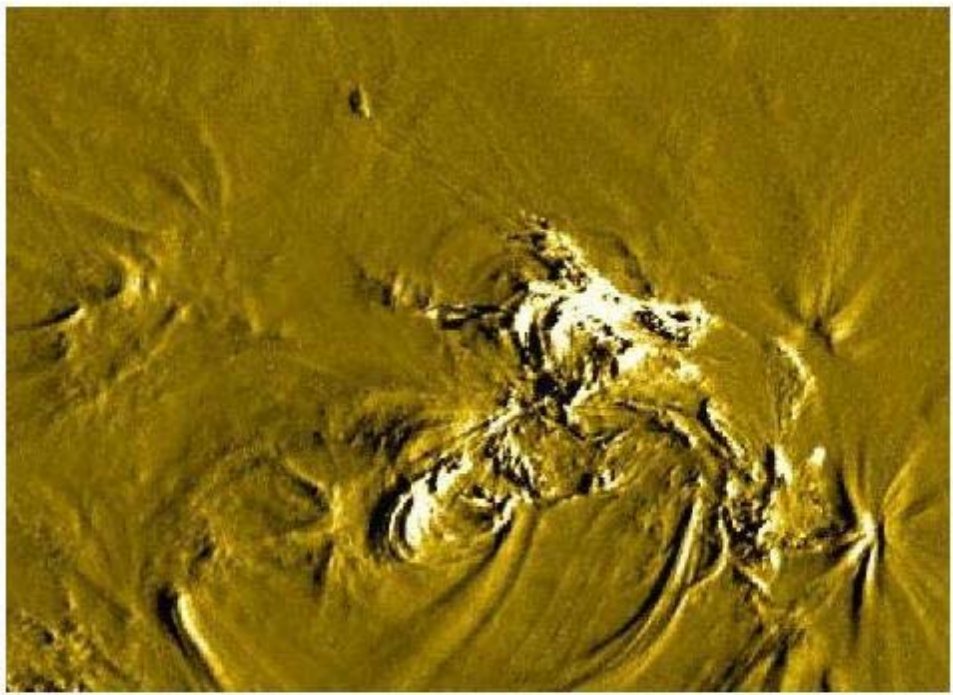

**FIGURE 15.** This is a "running difference" image of one small part of the sun's surface revealed by the TRACE satellite using a 171 Å filter. This filter is specifically sensitive to iron ion (Fe IX/X) emissions. The TRACE satellite recorded an eruption and mass ejection from this region of AR 9143 on 28 August 2000. A video of the flare event on 28 August 2000 is available here: http://vestige.lmsal.com/TRACE/Public/Gallery/Images/movies/T171_000828.avi

## ATOMIC MASS MEASUREMENTS

### The Source Of Luminosity In An Iron-Rich Sun

Figure 9 [53, 54, 76] is a summary of the events revealed by findings that *the Sun and its planets formed out of fresh supernova debris and inherited chemical and isotope heterogeneities from the parent star*:
   a.) Short-lived radioactive isotopes in the material that formed the planetary system (See Figures 1 and 2).
   b.) Isotope anomalies in meteorites, planets, and Sun from stellar nuclear reactions (See Figures 3, 4 and 7).
   c.) Element and isotope variations linked together in proto-planetary material (See Figure 8 and 10).

Evidence of mass-dependent fractionation, in the Sun and in the parent star that produced the solar system, was recorded by these separations of lightweight elements and isotopes from heavier ones:
   a.) Variations in the non-cosmogenic neon isotope abundances shown in Figures 5 and 6.
   b.) Linked fractionation-plus-nuclear effects in **FUN** inclusions of meteorites [44, 45] (See also Figure 4).
   c.) High abundances of lightweight isotopes in the solar wind (See Figure 12).
   d.) High abundances of lightweight elements in the photosphere (See Figures 11 and 13).
   e.) Images of a rigid, iron-rich structure below the Sun's fluid outer layers (See Figures 14 and 15).

The scenario in Figure 9 offers a viable solution for these observations, but that solution raises another puzzle: How can the Sun shine if its most abundant element is iron? The answer to this question, identified above as the *fourth enigma*, will also explain Fowler's *"solar neutrino problem"* [ref. 84, p. xi].

All iron isotopes have high nuclear stability [86]. Less abundant elements cannot explain solar luminosity. After correcting for fractionation, solar abundances of the elements correlate with nuclear stability – from the loosely bound nucleons of Li, Be and B to the tightly bound nucleons of Fe [87], as first suggested by Harkins [79].

In the search for an overlooked source of nuclear energy, students in an advanced nuclear chemistry course were assigned the task of re-examining the systematic properties of all 2,850 known nuclides (assemblages of neutrons, protons and electrons) [86]. Students were encouraged to abandon the conventional approach and to use reduced nuclear variables, like the reduced physical variables used in developing the corresponding states of gases [88].

When the nuclear charge, Z, and the mass number, A, are combined into one reduced variable, Z/A, the charge per nucleon, then all known nuclides lie in the range of $0 \geq Z/A \leq 1$. Likewise, the atomic mass and the mass number can be combined into the reduced variable used by Aston [3], M/A. Values of M/A, potential energy per nucleon, lie close to the value of 1.00 mass units per nucleon for all nuclides. Aston [3] subtracted 1.00 from the value of each to obtain a quantity called the *"packing fraction"* or *"nuclear packing fraction."*

The left side of Figure 16 shows the "Cradle of the Nuclides" [89] that is revealed when all 2,850 nuclides [86] are plotted in terms of these two reduced variables, Z/A and M/A, and then sorted by mass number, A.

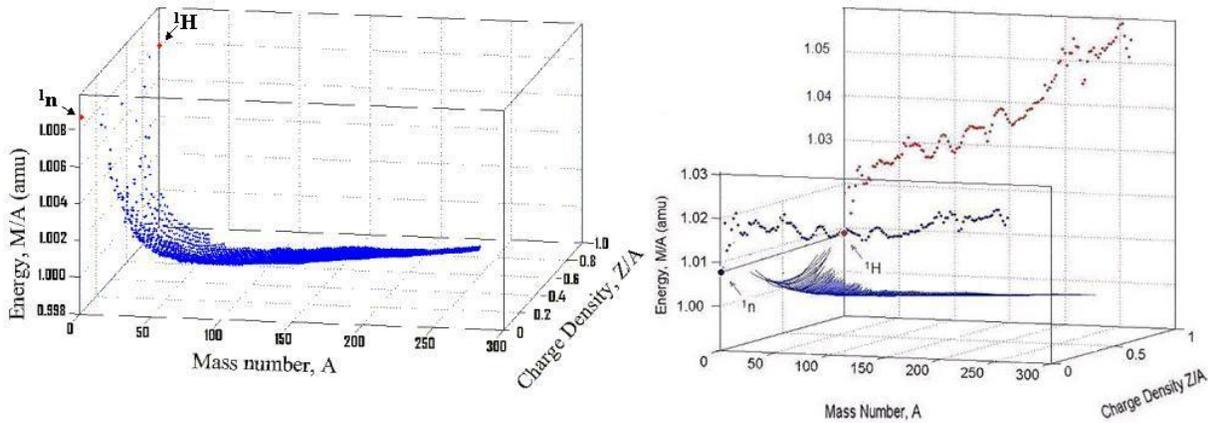

**FIGURE 16.** The "Cradle of the Nuclides" on the left [89] shows the potential energy per nucleon for all known nuclides, stable and radioactive nuclides [86]. The more stable nuclides have lower values of M/A and occupy lower positions in the cradle. Those that are radioactive or readily consumed by fusion or fission occupy higher positions. On the right, mass parabolas defined by the data points [86] intersect the front plane, at {Z/A = 0, M/A = (M/A)$_{neutron}$ + ~10 MeV}, for each value of A>1 [89-92].

The left side of Figure 16 shows all the atomic mass data [86]; the right side shows the mass parabolas defined by these data [86] at each value of A [86]. Intersections of mass parabolas with the front plane at Z/A = 0 and with the back plane at Z/A = 1.0, are shown on the right. The intercepts at Z/A = 0 and Z/A = 1.0 display peaks and valleys at the same mass numbers because of tight and loose nucleon packing, respectively. Additional repulsive interactions between positive charges cause the peaks to be more pronounced at Z/A = 1 than at Z/A =0 [90].

For purposes of explaining solar luminosity the most important revelation in Figure 16 is *repulsive interactions between like nucleons* (n-n or p-p) and attractive interactions between unlike nucleons (n-p) [93]. This suggests that *neutrons are in an excited state in neutron stars*, with about **10-22 MeV more energy** than a free neutron [89-91]. On the other hand, *the accepted view is that neutron stars are "dead" nuclear matter*, with the neutron at about **93 MeV less energy** than a free neutron [94].

In view of the above evidence for an iron-rich Sun that formed on the collapsed core of a supernova and recent SNO data for solar neutrinos [95], we conclude that *excited neutrons in the collapsed SN core continue to generate solar luminosity, solar neutrinos, and an outflow of 3 x 10$^{43}$ H$^\pm$ per year in the solar wind* [89-93, 96] by the following series of reaction:

- Neutron emission from the solar core          Generates >57% of solar luminosity
  $<_0^1n> \rightarrow {_0^1}n\ +\sim 10\text{-}22$ MeV
- Neutron decay          Generates < 5% of solar luminosity
  ${_0^1}n \rightarrow {_1^1}H^+ + e^- + \text{anti-}\nu\ + 0.78$ MeV
- Fusion and upward migration of H$^+$          Generates <38% of solar luminosity
  $4\ {_1^1}H^+ + 2\ e^- \rightarrow {_2^4}He^{++} + 2\ \nu\ + 27$ MeV
- Escape of excess H$^+$ in the solar wind          (~ 1% of the neutron-decay product)
  $3 \times 10^{43}$ H$^+$/yr $\rightarrow$ Departs in solar wind

The final section is a table summarizing our conclusions from puzzling isotope data and other unexpected observations since B2FH [14] explained the synthesis of elements in stars in their landmark 1957 paper:

# CONCLUSIONS

Table 1 shows the most obvious, common sense conclusions (middle column) to a seemingly complex set of observations (left column) made after B2FH [14] published their classical paper on element synthesis in stars [14].

**TABLE 1.** OBSERVATIONS AND COMCLUSIONS

| Observation/Questions/Enigmas | Conclusions (SN = Supernova) | Reference: Observation/Explanation |
|---|---|---|
| 1. Decay Products of Short-lived Isotopes | Fresh debris from a SN explosion 5 Gy ago formed the entire solar system | Figures 1 & 2, [4-6, 8-13, 17-19, 60]/ Figure 9, [17-18, 53-54, 73, 76, 80, 89] |
| 2. Isotope Anomalies in Stone Meteorites | The axial SN explosion left isotopes, elements unmixed in accretion disk | Figures 3, 4 & 7, [20-32, 43-52]/ Figure 9, [32, 46, 53-54, 73, 76, 80, 89] |
| 3. Isotope Anomalies in Iron Meteorites | Iron-rich SN debris directly formed iron meteorites and planetary cores | [55-59, 63-64]/ Figure 9, [53-54, 63-64, 80] |
| 4. Elements/Isotopes Were Linked **Xe-1** in FeS, **Xe-2** in Carbon Grains | **Xe-1** was made in iron-rich SN interior, **Xe-2** made near carbon-rich SN surface | Figures 8 & 10, [32, 53-54, 63-64, 67]/ Figure 9, [32, 46, 53-54, 73] |
| 5. Isotope Anomalies In Planets: **Xe-1** in Sun, Mars; **Xe-2** in Jupiter | **Xe-1** was made in iron-rich SN interior, **Xe-2** made near carbon-rich SN surface | Figures 8 & 10, [32, 53-54, 64, 67]/ Figure 9, [32, 46, 53-54, 73] |
| 6. Severely Mass-Fractionated Isotopes in Meteorites and Planets | Multi-stage mass separation in the Sun and in the parent star of the SN | Figures 4, 5, 6 &12, [20, 24, 33-41, 78]/ Figure 4, 5, 6 &12, [24, 33, 42, 78, 80] |
| 7. **FUN** (Fractionation + Nuclear) Effects Linked in Meteorites | The supernova made new isotopes in material that was mass fractionated | [44, 45]/ Figures 11 & 12, [44, 45, 78, 83] |
| 8. Mirror-Image Isotope Anomalies | Unmixed products of the various nuclear reactions that collectively made "normal" isotope abundances | Figure 7, [43-52] / Figure 9, [32, 46, 53-54, 73] |
| 9. P-1 Planetary Gas Component Had Only "Normal" **Xe-1**, Kr-1 and Ar-1 | This came from SN's iron-rich interior that was depleted of light elements | Figure 8, [32, 46, 53-54, 73, 76, 78]/ Figure 9, [32, 46, 53-54, 73, 76, 78] |
| 10. P-2 Planetary Gas Component Had "Strange" **Xe-2**, Kr-2, Ar-2, Ne, He | This came from the outer SN layers where light elements remained. | Figure 8, [32, 46, 53-54, 73, 76, 78]/ Figure 9, [32, 46, 53-54, 73, 76, 78] |
| 11. The Solar Surface Is Made Mostly of Light Elements | Elements undergo multi-stage mass separation in the Sun | Figure 11, [75]/ Figures 12 & 13 [78, 80, 83] |
| 12. Carbonaceous Meteorites Are Also Rich in Light Elements | These came mostly from the surface of the mass-fractionated parent star | Figure 11, [75]/ Figure 9, [32, 46, 53-54, 73] |
| 13. Why Does O/C ≈ 2 at the Surface of the Sun and Similar Stars? | Multi-stage mass separation decreases O/C ≈ 10 to O/C ≈ 2 at solar surface | Fowler [84]/ Figures 12 & 13 [78, 80, 83] |
| 14. What Are the Most Abundant Elements in the Solar System? | Iron, oxygen, nickel, silicon, sulfur, magnesium and calcium | [78-80, 87]/ Figure 9, [53, 54, 78-80, 87] |
| 15. What Causes the Solar Neutrino Deficit? | The number of neutrinos produced is the number detected. There is no deficit | Fowler [84]/ Figures 9, [90, 91, 92, 95, this paper] |
| 16. What Is The Source of Solar Luminosity? | Neutron emission and decay generate >62%; H-fusion generates <38% | Fowler [84]/ Figures 9, [90, 91, 92, 95, this paper] |
| 17. What Is the Source of Hydrogen in the Solar and Stellar Winds? | Neutron-decay and upward acceleration of $H^+$ ions by solar magnetic fields | [89-93, 96] / [96] |

# ACKNOWLEDGMENTS

Support from the University of Missouri-Rolla and the Foundation for Chemical Research, Inc. (FCR) and permission to reproduce figures from reports to FCR are gratefully acknowledged. NASA and ESA, specifically the SOHO and TRACE programs, made possible the solar images shown in Figures 14 and 15. Students enrolled in Advanced Nuclear Chemistry (Chem. 471) in the spring semester of 2000 – Cynthia Bolon, Shelonda Finch, Daniel Ragland, Matthew Seelke and Bing Zhang – contributed to the development of the "Cradle of the Nuclides" (Figure 16) that exposed repulsive interactions between neutrons [89-92]. This manuscript benefited from comments by UMR Chancellor Gary Thomas, Eugene Savov (Bulgarian Academy of Sciences), Kiril Panov (Director, Institute of Astronomy, Bulgarian Academy of Sciences), Michael Ibison (Institute for Advanced Studies, Austin), José B. de Almeida (University of Minho), R. Greg Downing (NIST), Ramachandran Ganapathy (Baker Chemical Co.), Hilton Ratcliffe (South Africa Astronomical Society), and Bing Zhang (GE Global Research, Shanghai). This paper is dedicated to the memory of Dr. Dwarka Das Sabu who participated in many of the findings [28, 32, 42, 53, 54, 65] that laid the basis for the conclusions reached here.

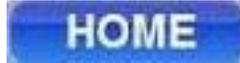